%% file: EUVIP_2022.tex
\documentclass[conference]{IEEEtran}
\IEEEoverridecommandlockouts 
\usepackage{amsmath,graphicx}
\usepackage{cite}
\usepackage{balance}
\usepackage{multirow}
\usepackage{makecell}
\usepackage{pgfplots}
\usepackage{amssymb,amsfonts}
\usepackage{algorithmic}
\usepackage{xcolor}
\usepackage{boldline, soul}
\usepackage{booktabs}
\usepackage{indentfirst}
\let\labelindent\relax
\usepackage{enumitem}
\usepackage{pifont}
\usepackage{pdfpages}
\usepackage{textcomp}
\usepackage[acronyms]{glossaries}
\usepackage{algorithm}
\usepackage{algorithmic}
\usepackage{caption}
\usepackage{mathtools}
\usepackage{pdfpages}
\usepackage{subcaption}
\usepackage{amsmath}
\DeclareMathOperator*{\argmaxA}{arg\,max} 

\newcolumntype{P}[1]{>{\centering\arraybackslash}p{#1}}
\newcolumntype{M}[1]{>{\centering\arraybackslash}m{#1}}
\newcolumntype{C}[1]{>{\raggedright\arraybackslash}m{#1}}

\def\BibTeX{{\rm B\kern-.05em{\sc i\kern-.025em b}\kern-.08em
    T\kern-.1667em\lower.7ex\hbox{E}\kern-.125emX}}
    
    \setcellgapes{3pt}
\def\BibTeX{{\rm B\kern-.05em{\sc i\kern-.025em b}\kern-.08em
    T\kern-.1667em\lower.7ex\hbox{E}\kern-.125emX}}

\pgfplotsset{compat=1.13}

\usepackage{hyperref}
\hypersetup{
    colorlinks=true,
    linkcolor=blue,
    filecolor=magenta,      
    urlcolor=cyan,
}

\input{acronym}

\begin{document}

\title{Ensemble Learning for Efficient VVC Bitrate Ladder Prediction}
\author{

\IEEEauthorblockN{Fatemeh~Nasiri$^{1, 2, 3}$, Wassim Hamidouche$^{1,2}$, Luce Morin$^{1, 2}$, Nicolas Dholland$^{3}$ and Jean-Yves~Aubi\'{e}$^{1}$}

\IEEEauthorblockA{$^1$ IRT b$<>$com, 35510 Cesson-S\'{e}vign\'{e}, France,\\
		 $^2$ Univ Rennes, INSA Rennes, CNRS, IETR - UMR 6164, 35000 Rennes, France \\
		 $^3$ AVIWEST-Haivision, 35760, Saint-Gr\'{e}goire, France
		 } 
}

\maketitle

\begin{abstract}
Changing the encoding parameters, in particular the video resolution, is a common practice before transcoding. To this end, streaming and broadcast platforms benefit from so-called bitrate ladders to determine the optimal resolution for given bitrates. However, the task of determining the bitrate ladder can usually be challenging as, on one hand, so-called fit-for-all static ladders would waste bandwidth, and on the other hand, fully specialized ladders are often not affordable in terms of computational complexity. In this paper, we propose an ML-based scheme for predicting the bitrate ladder based on the content of the video. The baseline of our solution predicts the bitrate ladder using two constituent methods, which require no encoding passes. To further enhance the performance of the constituent methods, we integrate a conditional ensemble method to aggregate their decisions, with a negligibly limited number of encoding passes. The experiment, carried out on the optimized software encoder implementation of the VVC standard, called VVenC, shows significant performance improvement. When compared to static bitrate ladder, the proposed method can offer about 13\% bitrate reduction in terms of BD-BR with a negligible additional computational overhead. Conversely, when compared to the fully specialized bitrate ladder method, the proposed method can offer about 86\% to 92\% complexity reduction, at cost the of only 0.8\% to 0.9\% coding efficiency drop in terms of BD-BR.
\end{abstract}


\section{Introduction}
Network heterogeneity, varying users' display size, and various video contents with different spatio-temporal features are all factors that could impact the performance of live video streaming or \gls{VOD} services. As a result, \gls{dash}  \cite{sodagar2011mpeg} and \gls{HLS} \cite{HLS} are two main industrial technologies that have been widely adopted in the media industry to incorporate heterogeneous network conditions. In both technologies, the input video is potentially down-sampled from its native resolution changes before encoding, in order to meet the available constraints such as bandwidth, complexity and latency. 

\begin{figure}[!ht]
    \centering
    \input{figs/RD_firstPage}
    
    \caption{Bitrate points for switching between resolutions for two sequences with different spatio-temporal characteristics. Left: complex motion and texture, right: simple motion and texture.}
    \label{fig:RD_firstpage1}
\end{figure}
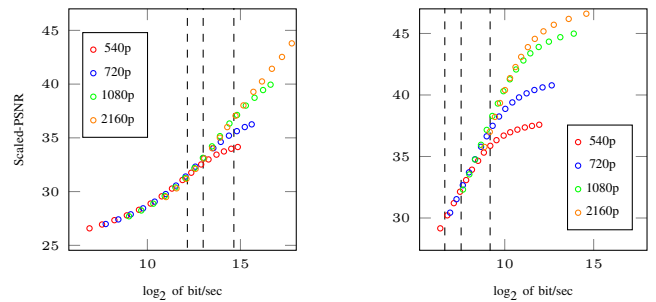

The traditional approach to change the resolution is performed by employing the so-called ``bitrate ladder'' \cite{kaafarani2021evaluation}. A bitrate ladder recommends the resolution for a given bitrate, by dividing the bitrate range into a set of predefined bitrate intervals and associating ascending resolutions to consecutive intervals. The simplest implementation of this idea is called static bitrate ladder, where one ladder is fit for all types of video contents. The main drawback of a static bitrate ladder is that its recommendation scheme is the same for all video contents, regardless of their spatio-temporal features. To elaborate this shortcoming, Fig. \ref{fig:RD_firstpage1} shows how the optimal points for changing from one resolution to another might vary, depending on spatio-temporal features.

There are mainly two categories of solution for bitrate ladder prediction. The first category relies on an exhaustive encoding, while proposing to accelerate the encoding process to make their complexity affordable. In the work presented in \cite{goswami2018adaptive}, the coding information extracted from encoding in the lowest resolution are used to speed up the encoding process at higher resolutions. In this work, the coding decisions including \gls{cu} quad-tree structure and \gls{pu} predictions, coding modes and \glspl{mv} information are derived from low resolution video to reduce the overall number of \gls{rdo} calculation at higher resolutions. Furthermore, Brightcove proposes a multi-codec bitrate ladder in which two codecs including \gls{avc} and \gls{hevc} are considered to be used by clients \cite{reznik2019optimal}. Finally, in \cite{ccetinkaya2021fast}, an \gls{ann}-based approach is used for a fast multi-resolution and multi-rate encoding. For multi-rate encoding, the lowest bitrate representation and for multi-resolution encoding, the highest bitrate from the lowest resolution representation are chosen as the reference, respectively. Then the \gls{ctu} split decisions are predicted by using the pixel values from the target resolution and encoding information from the reference representation.

In the second category of solutions, the additional encodings are partially or entirely replaced by methods that directly or indirectly predict the ladder. In one of the simplest realization of this category, the work presented in \cite{lederer2012dynamic} trains separate bitrate ladders for different pre-defined categories of video contents. As a result, each new video has to be first classified, then adopt one of the trained ladders. In another solution proposed by Bitmovin \cite{bitmovin}, first, a variety of features such as frame rate, resolution and resulting bitrate from multiple encodings is extracted from the source video. Then, a \gls{ml}-based method is used to predict the convex hull and adjust an optimized profile for encoding the video. Likewise, Cambria \cite{Cambria} proposes a method named Source Adaptive Bitrate Ladder (SABL). They run a fast \gls{CRF} encoding to estimate the encoding complexity. The obtained results from this encoding are then used to adjust the encoding ladder up or down. Moreover, MUX \cite{mux} proposes a neural network based solution for estimating the bitrate ladder which the new videos loaded into the network are contributed back to the training set. Furthermore, the work presented in \cite{katsenou2021efficient} introduces a method to predict the \glspl{qp} of the crossover points between the RD curves of two consecutive resolutions, that finally construct the bitrate ladder by performing several encodings in the predicted crossover \glspl{qp}. 
In the work of \cite{toni2015optimal}, the bitrate ladder identification problem was solved using integer linear programming, while maximizing \gls{qoe} measured using National Telecommunications and Information Administration Video Quality Metrics (NTIA VQM)~\cite{sector2008objective}. 

In this work, the prediction of the bitrate ladder is based on an ensemble learning method that aggregates the decision of two constituent \gls{ml}-based methods. If necessary, the proposed aggregator might conduct limited additional encodings to make the final decision about the optimal ladder. The two \gls{ml} methods are trained by the low-level features extracted from the video in its native resolution and the corresponding bitrate-quality-resolution points. 

The remaining of the paper is organized as follows. Section \ref{sec:Problem definition} formulates the problem definition of bitrate ladder prediction, while Section~\ref{sec:pro} explains the proposed ML-based method. The experimental results and discussions showing the coding efficiency of the proposed method are presented in Section~\ref{sec:res}, and finally, Section~\ref{sec:conc} concludes this paper.

\section{Problem Formulation}
\label{sec:Problem definition}

Let $v$ be an input video sequence and $S=\{s_1, s_2, ..., s_{|S|}\}$ a set  of resolutions in which $v$ can be encoded. An encoder is also given whose task can be simplified in a function, denoted as $E$, which receives $v$ and a resolution $s_i \in S$, as well as a target bitrate $r$. The simplified output of this encoder is a quality index $q$. Without loss of generality, we assume that the quality metric can potentially be any of the common objective metrics such as \gls{psnr}, \gls{vmaf} or \gls{msssim}.

Encoding a video sequence $v$ at resolution $s_i$ and bitrate $r$ with an output quality $q$ can be expressed as:
\begin{equation}
  q =  E(v,r,s_i), \hspace{13px} \text{ where } s_i \in S.
\end{equation}

\begin{figure}[!ht]
    \centering
    \input{figs/RD_convex_hull_ladder}
    
    \caption{Four stages of constructing the bitrate ladder (d) from the full rate-quality points (a), through the convex-hull (b) and cross-over bitrate computations (c).}
    \label{fig:RD_convex_hull_ladder}
\end{figure}
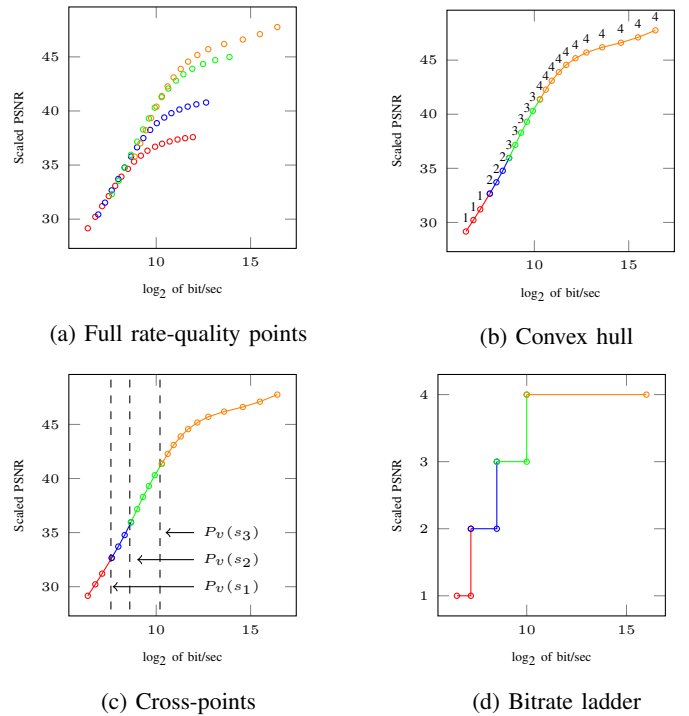

For a video sequence $v$, by varying the two parameters $r$ and $s_i$, the output qualities of encoder $E$ generate a diagram of full rate-quality operating points, as shown in Fig. \ref{fig:RD_convex_hull_ladder}-(a). This diagram is used as the starting point for the task of bitrate ladder prediction. 

Given a full rate-quality operating point diagram, the convex hull of a video $v$ can be expressed as a function of rate $r$ as follows:

\begin{equation}
\label{eq:convexhull}
\begin{split}
    q^* = C_v(r)  & \text{ where } E(v,r,s_i) \leq q^*\\ 
    & \text{ for all } s_i \in S
\end{split}
\end{equation}

In other words, the convex hull function $C_v(r)$ determines the highest quality that can be obtained for a video $v$ after encoding with $E$ in the available resolutions $S$. This function has been visualized in Fig. \ref{fig:RD_convex_hull_ladder}-(b), where labels and colorization at given bitrate points indicate the resolution that is resulting in the optimal quality $q^*$.

In this work, we assume that convex hulls are monotonic, and moreover, each resolution switch is imperatively from resolution $s_i$ (where $1\leq i < |s|$) to resolution $s_{i+1}$, which is the immediate next larger available resolution. Therefore, for each resolution $s_i$, a bitrate point can be calculated in which the resolution switch must be applied. This bitrate is called the cross-over bitrate of resolution $s_i$ in the rest of this paper and is expressed as:

\begin{equation}
\label{eq:crosspoint}
\begin{split}
    r^*_i & = P_v(s_i)   \text{ where } \\
     C_v(r^*_i) & =E(v,r_i^*,s_i) \text{ and}\\
     C_v(r^*_i+\epsilon) & =E(v,r_i^*+\epsilon,s_{i+1}).
\end{split}
\end{equation}

In other words, Eq. \eqref{eq:crosspoint} computes for a given resolution $s_i$, the largest bitrate point as $r^*_i$, where the highest quality $q^*$ is obtained by encoding in resolution $s_i$. While after that point (\textit{i.e.} addition of $\epsilon$, where $\epsilon>0$), a resolution switch to $s_{i+1}$ is needed. Fig. \ref{fig:RD_convex_hull_ladder}-(c) demonstrates an example computation of cross-over bitrates. 

The bitrate ladder of a sequence is defined as a function that determines the optimal resolution for any given bitrate. A trivial approach to compute the bitrate ladder of sequence is to actually encode it in all available resolutions and sufficient number of bitrates. By doing so, one can obtain the full rate-quality operating points needed for Eq. \eqref{eq:convexhull}  and Eq. \eqref{eq:crosspoint}. At this point, the reference bitrate ladder of video $v$ in resolutions defined in $S$, can be expressed as in Eq. \eqref{eq:ladder}. Fig. \ref{fig:RD_convex_hull_ladder}-d visualizes an example of reference bitrate ladder computed from all operational rate-quality points. 

\begin{equation}
\label{eq:ladder}
    \begin{split}
        i^* = L^*_{v,S}(r) \text{ where } P_v(s_{i-1}) < r \leq P_v(s_i).
    \end{split}
\end{equation}

In this paper, a \gls{ml}-based method is used to learn how to construct bitrate ladder of a video sequence, without having to encode it in all resolutions:

\begin{equation}
    \hat{L}_{v,S} = F(v,S), 
\end{equation}


\section{Proposed ensemble bitrate ladder prediction}
\label{sec:pro}
\subsection{Framework}

The main contribution of this paper is the deployment of ensemble machine learning model, which is a mechanism that allows combining multiple predictions coming from its constituent learning algorithms. The number of constituent methods can vary from two to several methods depending on the performance of the methods. The prediction process and inputs can be different in each \gls{ml} method, however, eventually the best resolution for a given bitrate is the output. In the proposed framework, we use an ensemble aggregator method to collect the output of all constituent methods and provide the final bitrate ladder. 

\begin{figure*}
    \centering
    \includegraphics[width=0.8\textwidth, angle=0]{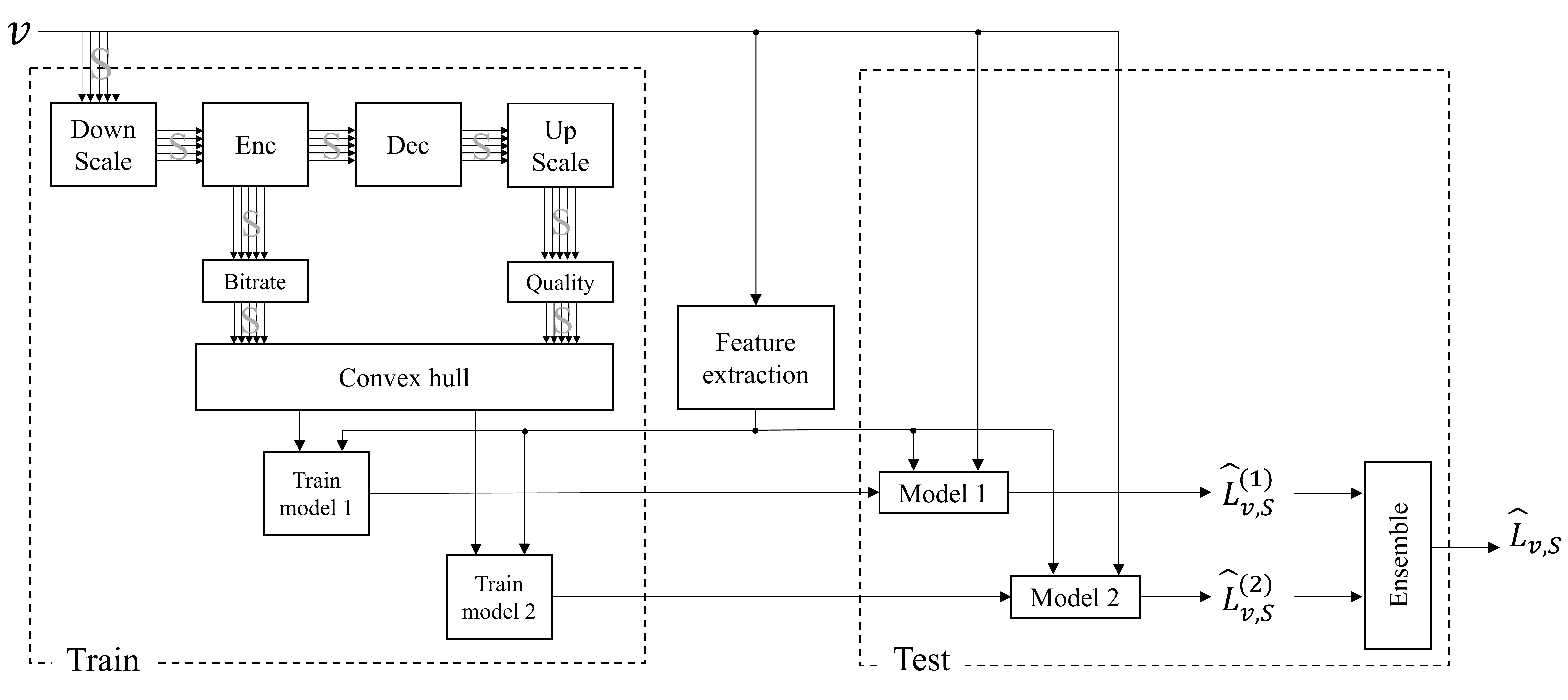}
    \caption{Framework of proposed method, including the "train" and "test" phases. The parallel arrows indicate the process has been carried out in all available resolutions of $S$. }
    \label{fig:fr_main}
\end{figure*}

Fig.~\ref{fig:fr_main} shows the overall framework of our proposed method, including two main phases of ``train'' and ``test''. These two phases share a feature extraction step, which serves for the training and testing of the two constituent bitrate ladder prediction methods. The input video(s) $v$ is to be represented in the highest possible resolution, specified by $S$. In the train phase, the goal is to independently train the two constituent methods, such that they can individually predict the bitrate ladder for any given video in the test phase. To do so, the high resolution input is down-sampled, encoded, decoded and finally up-sampled, in order to provide the bitrate-quality points needed to construct the ground-truth bitrate ladder. In the test phase, the two constituent methods are used to predict two potentially different ladders, which are then used as inputs to the ensemble aggregator for producing the final bitrate ladder prediction.

\subsection{Classifier constituent predictor}

As the first constituent bitrate ladder prediction method, a multi-class classifier is used. At the core of this method, model $M^{Cl}$ is trained that receives as input, the video sequence $v$ and the target bitrate $r$, while the output is the index of predicted optimal resolution, defined in $S$: 

\begin{equation}
    \hat{i} = M^{Cl}_S(v,r).
\end{equation}

In other words, the first method directly predicts the value $i$ in Eq. \eqref{eq:ladder}, without having to compute the cross-over points $P_v$, denoted in Eq. \eqref{eq:crosspoint}. Therefore, by applying the core model $M_S^{Cl}$ to all bitrate values, one can express the global operation of the classifier constituent predictor as:

\begin{equation}
    \hat{L}_{v,S}^{Cl}  = F^{Cl}(v,S).  
\end{equation}

\subsection{Regressor constituent predictor}
In the second method, a regressor is used to predict the cross-over bitrate points. Given a resolution $s_i$ (where $1 \leq i < |S|$) of the video sequence $v$, the regressor model that has learned the operation in Eq. \eqref{eq:crosspoint}, predicts at which bitrate the resolution should be switched to $s_{i+1}$:

\begin{equation}
    \hat{r}_i = M^{Rg}_S(v,s_i).
\end{equation}

By applying the regressor model in Eq. \eqref{eq:ladder}, to identify the cross-over bitrates, one can express the second constituent predictor as:

\begin{equation}
     \hat{L}_{v,S}^{Rg} = F^{Rg}(v,S)
\end{equation}

\subsection{Ensemble aggregator}
Once the two predictions of the bitrate ladder are computed by the constituent methods, the ensemble aggregator combines the two ladders and produces the final output, as:
\begin{equation}
\label{eq:aggregator}
    \hat{L}_{v,S}  = Agr(F^{Cl}, F^{Rg}) = F(v,S).
\end{equation}

Algorithm \ref{algorithm1} describes how the function $Agr$ in Eq. \eqref{eq:aggregator} computes the final predicted bitrate ladder. The goal of this function is to take into account the two predictions made by the two constituents and determine the final resolution for each bitrate point. In case that the two  constituent predictions are the same, the aggregation is simply done by choosing the common prediction. However, in case of different predictions, additional encodings by $E$ are carried out to make the final decision. The number of encodings depends on a parameter, denoted as $isFast$ in  Algorithm \ref{algorithm1}. If the fast mode is used, encoding is carried out only with the two resolutions predicted by the constituent methods. Otherwise, all possible resolutions are tested. In contrast with the ``fast'' mode, this mode is called the ``full'' mode in the rest of this paper. Either mode, the resolution that provides the highest quality among the tested encodings is selected.

\input{figs/algorithm}

\subsection{Training process}
\subsubsection{Dataset}
One of the crucial steps in \gls{ml} based methods is to have a large number of sequences for training the models. Therefore, we gathered a dataset of 100 videos from public and private sources including: BVI SR \cite{mackin2018study}, Derf collection \cite{derf}, MCML \cite{cheon2017subjective}, SJTU \cite{song2013sjtu} and UGV \cite{mercat2020uvg}. All sequences have the native resolution of 3840$\times$2160p with the frame rate of 60 fps. We have converted sequences in 10 bits to 8 bits and all the other color formats to 4:2:0 format. As the sequences have different duration, they have been split into chunks of one second (64 frames). It is worth mentioning that an additional scene change detection has been applied in order to ensure content homogeneity within each chunk and content diversity between different chunks.   


\begin{figure}[!ht]
    \centering
    \input{figs/SI_TI_scatter}
    \label{fig:sitiscatter}
\end{figure}

In order to show the diversity of the dataset, we have computed \gls{si} and \gls{ti} descriptors \cite{winkler2012analysis}. In Fig. \ref{fig:sitiscatter}, the distribution of these two spatial and temporal descriptors are shown. As can be seen, the selected dataset covers a wide range of the spatial and temporal characteristics.     

\subsubsection{Features}
The videos with a complex spatial characteristics are likely to have larger difference between neighboring pixels. Thus, in this work, we use \gls{glcm} \cite{haralick1973textural} which is a traditional spatial features descriptor and has been used in many studies for demonstrating the spatial complexity. \gls{glcm} is composed of intensity contrast of neighboring pixels in a video frame. Therefore, we can capture the level of coarseness as well as directional information of the video texture. \gls{glcm} has five main descriptors: contrast, correlation, energy, homogeneity and entropy. In addition, to capture the temporal characteristics of the video, we have extracted the \gls{TC} from two consecutive frames through the frames of the video. Prior to using these features to predict the bitrate ladder, we have used the recursive feature elimination method \cite{kuhn2013applied} to select the most effective features. 

\subsection{ML methods}
In order to find the proper ML methods for regression and classification, we trained and tested several methods. For classification, the decision tree classifier with gradient boost methods provided the best result compared to other kernels. Similarly, for the regressor models, after testing several methods, \gls{GP} provided the best results compared to other methods. Thus, we used the \gls{GP} as the regressor for predicting the three cross-over bitrates. 


%
\section{Experimental results}
\label{sec:res}
\subsection{Experiment setting}

In the experiment, four resolutions are employed such that $S=\{2160p, 1080p, 720p, 540p\}$. For down-scaling the video sequences, the Ffmpeg~\cite{ffmpeg} implementation of the Lanczos filter \cite{duchon1979lanczos} has been used. In order to upscale the videos, we use the same filter in FFMPEG to bring back the down-scaled videos into their native resolution. As all \gls{psnr} computations are computed in the native resolution (\textit{e.g.} 2160p), the scaled \gls{psnr} metric has been used \cite{helmrich2020study}.  

The VVC codec that has been used is the latest version of VVenC \cite{wieckowski2021vvenc}, in the ``faster'' quality preset. As VVC has not been widely used in any sector of the streaming/broadcast ecosystem, there is neither officially nor unofficially no defined static \gls{vvc} bitrate ladder in the literature/industry. In order to address this issue and provide a reference point to our performance measurements, we calculated the average bitrate ladder through our training dataset and considered it as the static \gls{vvc} bitrate ladder in the experiments. In addition to the static ladder, the fully specialized bitrate ladders computed from exhaustive encoding in different resolutions for each sequence in the dataset have also been used as benchmark. This ladder is referred to as the \gls{gt} ladder in the results section.

For the evaluation, different metrics such as \gls{bd-br} \cite{bjontegaard2001calculation} and prediction accuracy are used. For comparing the bitrate ladders, we constructed R-D curves of available rate and distortions values and compared them with \gls{bd-br} metrics. To compute the \gls{bd-br} metric given two bitrate ladders, one ladder is chosen as the ``reference'', while the other one as the ``test''. Video sequences are then encoded in several bitrates, while their resolution is determined once by the ``reference'' ladder and once by the ``test'' ladder. The bitrate and scaled \gls{psnr} values are then collected and used with a mildly modified \gls{bd-br} computation in order to enable it with more than four operational bitrate-quality points. Finally, in order to avoid over-fitting, the results are the output of tenfold cross-validation, and all the metrics are averaged over the ten folds.

\subsection{Results}
Table \ref{tab:results} summarizes the coding efficiency evaluation of different settings of the proposed method. Notably, the first two rows present the performance of the two constituent predictors, when used outside the proposed ensemble framework. The last two rows are consequently the proposed ensemble method, when the ``fast'' and ``full'' modes are used, respectively. 


The first metric demonstrates the accuracy of each method in exact prediction of the optimal resolution over all tested bitrates. While the second and third metrics indicate the \gls{bd-br} performance versus the \gls{gt} and static bitrate ladders, respectively. It is noteworthy that the negative values of the \gls{bd-br} metric indicate bitrate saving in the same level of quality, hence, should be considered as improvement of performance.

The first observation is that the regressor method globally has a better performance than the classification method. However, both ensemble methods (with fast and full encoding) outperform the regressor method, in all three metrics. This proves that the ensemble approach is indeed helping the grasp the best out of each constituent predictor.

\begin{table}[!ht]
    \caption{Average performance metrics of four different versions of the proposed method.}
    \centering
    \input{figs/average_res}
    \label{tab:results}
\end{table}

Fig. \ref{tab:results_all} provides a more detailed view on the \gls{bd-br} performance. Each diagram in this figure presents histogram of \gls{bd-br} metric on the test sequences. At left, the \gls{gt} ladder has been used as reference and positive \gls{bd-br} values indicate bitrate increase. Hence, being smaller is better.
In this sense, both ensemble methods significantly outperform the classification and regressor methods. Inversely, the results presented at right are obtained by using the static bitrate ladder as reference. Hence, more negative values means more gain.   

\begin{figure}[!ht]
    \centering
    \input{figs/hist_sample}
   
\caption{Distribution of \gls{bd-br} on test sequences}
 \label{tab:results_all}
\end{figure}
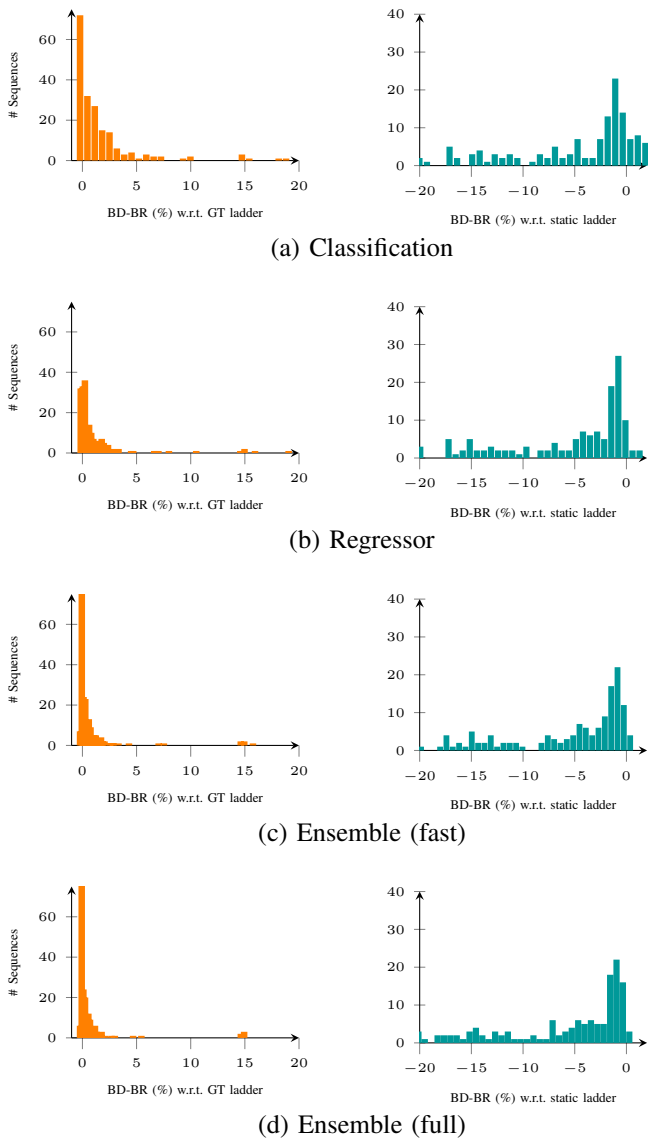

\begin{figure}[!ht]
    \centering
    \input{figs/comp_vs_bdbr}
    
\caption{\gls{bd-br} vs. complexity evolution of different methods. The numbers in parenthesis indicate the overhead in terms of encoding time with respect to the \gls{gt} method as reference. }
\label{tab:validationReg}
\end{figure}
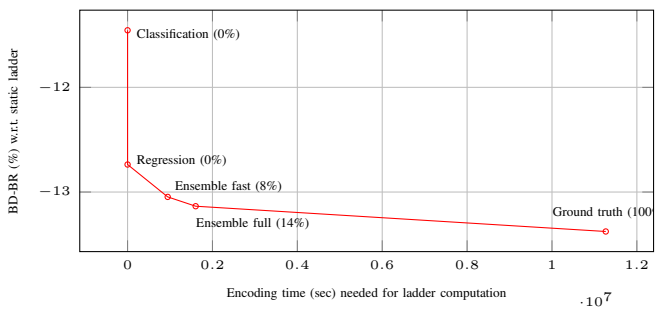

The additional gain brought by the ensemble methods is at the cost of encodings needed to aggregate decisions. To understand this impact, Fig. \ref{tab:validationReg} demonstrates 
the average bitrate gain compared to the static bitrate ladder of different methods with respect to their complexity. The complexity metric of this experiment was the total encoding time spent for generating necessary bitrate-quality points of each method. As shown, the \gls{gt} bitrate ladder method is highly complexity-intensive, while a significant portion of its \gls{bd-br} gain can be achieved by the proposed methods at much lower complexity. Conversely on the low complexity extreme of the diagram, the two methods of classification and regressor impose no complexity overhead. However, their performance can be noticeably improved with a limited number of additional encodings. 

\section{Conclusion}
\label{sec:conc}
This paper proposes an ML-based method for predicting the bitrate ladder in adaptive streaming use-cases. The proposed method fits two supervised machine learning methods on a set of spatio-temporal features extracted from each sequence, in order to learn their ground truth bitrate ladder. An ensemble aggregation method is then used to improve the performance of the two constituent methods at the cost of additional encodings. The performance of the proposed solution is assessed using a static and fully customized ground-truth bitrate ladders as benchmark methods. Compared to static ladder, the proposed method is able to achieve 13\% coding efficiency gain in terms of \gls{bd-br}, with negligible added complexity. While, compared to the fully customized ladder, the proposed method can significantly reduce the complexity at the cost of only 0.77\% \gls{bd-br} coding efficiency loss.


\balance
\bibliographystyle{unsrt}
\bibliography{mybib}

\end{document}

%% file: acronym.tex
\makeglossaries
\newacronym{ml}{ML}{Machine Learning}
\newacronym{cf}{CF}{Colourfulness}
\newacronym{si}{SI}{Spatial information}
\newacronym{ti}{TI}{Temporal information}

\newacronym{cnn}{CNN}{Convolutional Neural Network}
\newacronym{ann}{ANN}{Artificial Neural Network}
\newacronym{vvc}{VVC}{Versatile Video Coding}
\newacronym{hevc}{HEVC}{High Efficiency Video Coding}
\newacronym{dbf}{DBF}{De-Blocking filter}
\newacronym{qp}{QP}{Quantization Parameter}
\newacronym{sao}{SAO}{Sample Adaptive Offset}
\newacronym{alf}{ALF}{Adaptive Loop Filter}
\newacronym{sr}{SR}{Super Resolution}
\newacronym{pp}{PP}{Post Processing}
\newacronym{ilf}{ILF}{In-Loop Filtering}
\newacronym{qe}{QE}{Quality Enhancement}
\newacronym{gop}{GoP}{Group of Pictures}
\newacronym{evc}{EVC}{Essential Video Coding}
\newacronym{ctu}{CTU}{Coding Tree Unit}
\newacronym{mse}{MSE}{Mean Squared Error}
\newacronym{cu}{CU}{Coding Unit}
\newacronym{pu}{PU}{Prediction Unit}
\newacronym{ipm}{IPM} {Intra Prediction Modes}
\newacronym{tu}{TU}{Transform Unit}
\newacronym{msssim}{MS-SSIM}{Multi-Scale Structural SIMilarity}
\newacronym{pqf}{PQF}{Peak Quality Frame}
\newacronym{mc}{MC}{Motion Compensated}
\newacronym{gpu}{GPU}{Graphic Processing Unit}
\newacronym{ai}{AI}{All Intra}
\newacronym{ra}{RA}{Random Access}
\newacronym{ld}{LD}{Low Delay}
\newacronym{vr}{VR}{Virtual Reality}
\newacronym{dc}{DC}{direct current}
\newacronym{bd-br}{BD-BR}{Bj\o ntegaard Delta Bit Rate}
\newacronym{ct}{CT}{Coding Type} 
\newacronym{rt}{RT}{Run-Time}
\newacronym{rdo}{RDO}{Rate Distortion Optimization}
\newacronym{avc}{AVC}{Advanced Video Coding}

\newacronym{qoe}{QoE}{Quality of Experience}

\newacronym{mv}{MV}{Motion Vector}
\newacronym{psnr}{PSNR}{Peak Signal-to-Noise Ratio}
\newacronym{vmaf}{VMAF}{Video Multi-Method Assessment Fusion}

\newacronym{mae}{MAE}{Mean Absolute Error}

\newacronym{ctc}{CTC}{Common Test Condition} 
\newacronym{tum}{TUM}{Transform Units Map}
\newacronym{pm}{PM}{Partitioning Map}
\newacronym{qm}{QM}{QP Map}
\newacronym{cum}{CUM}{Coding Units Map}
\newacronym{resm}{ResM}{Residual Map}
\newacronym{predm}{PrM}{Prediction Map}
\newacronym{mm}{MM}{Mean Map}
\newacronym{ipmm}{IPMM}{Intra Prediction Mode Map}

\newacronym{dcad}{DCAD}{Deep CNN-based Auto Decoder}
\newacronym{IFCNN}{IFCNN}{In-Loop filter CNN }
\newacronym{STResNet}{STResNet}{Spatial-Temporal Residue Network}
\newacronym{DCAD}{DCAD}{Deep CNN-based Auto Encoder}
\newacronym{DSCNN}{DSCNN}{Decoder-side Scalable Convolutional Neural Network}
\newacronym{MMS-net}{MMS-net}{Multi-modal/Multi-scale Convolutional Neural Network}
\newacronym{VRCNN}{VRCNN}{Variable-filter-size Residue-learning CNN}
\newacronym{MSDD}{MSDD}{Multi-Scale deep decoder}
\newacronym{QECNN}{QECNN}{Quality Enhancement Convolutional Neural Network}
\newacronym{MFQE}{MFQE}{Multi-Frame Quality Enhancement }
\newacronym{CNNF}{CNNF}{CNN Filter}
\newacronym{RHCNN}{RHCNN}{Residual Highway CNN}
\newacronym{FECNN}{FECNN}{Frame Enhancement CNN}
\newacronym{Residual-VRN}{Residual-VRN}{Residual-based Video Restoration Network}
\newacronym{MGANet}{MGANet}{Multi-frame Guided Attention Network}
\newacronym{MLSDRN}{MLSDRN}{Multi-Channel Long-Short term Dependency Residual Network}
\newacronym{ADCNN}{ADCNN}{Attention based Dual-scale CNN}
\newacronym{MM-CU-PRN}{MM-CU-PRN}{Multi-scale Mean value of CU-Progressive Rethinking Network}
\newacronym{SDTS}{SDTS}{Spatial Details and Temporal Structure}
\newacronym{VRCNN-BN}{VRCNN-BN}{Variable Filter-size Residue-learning CNN with batch normalization layer}
\newacronym{MIF}{MIF}{Multi-frame In-loop Filter}
\newacronym{MRRN}{MRRN}{Multi-Reconstruction Recurrent Residual Network}
\newacronym{RRCNN}{RRCNN}{Recursive Residual Convolution Neural Network}
\newacronym{B-DRRN}{B-DRRN}{Deep Recursive Residual Network with Block information}
\newacronym{CPHER}{CPHER}{Coding Prior based High Efficiency Restoration}
\newacronym{MPRN}{MPRN}{Multi-level Progressive Refinement Network}
\newacronym{BDRRN}{BDRRN}{Block Information Constrained Deep Recursive Residual Network}
\newacronym{SEFCNN}{SEFCNN}{Squeeze-and-Excitation Filtering CNN}
\newacronym{GAN}{GAN}{Generative Adversarial Network}
\newacronym{MFRNet}{MFRNet}{Multi-level Feature review Residual dense Network}
\newacronym{IResLB}{IResLB}{Inception and Residual Learning based Block}

\newacronym{tcp}{TCP}{Transmission Control Protocol}
\newacronym{ott}{OTT}{over-the-top}
\newacronym{has}{HAS}{HTTP Adaptive Streaming}
\newacronym{udp}{UDP}{User Datagram Protocol}
\newacronym{nat}{NAT}{Network address translation}
\newacronym{abr}{ABR}{Adaptive Bitrate Selection}
\newacronym{cdn}{CDN}{Content delivery network}
\newacronym{mpd}{MPD}{Media Presentation Description}
\newacronym{uri}{URI}{Uniform Resource Identifier}
\newacronym{MiPSO}{MiPSO}{Multi-Period Per-Scene Optimization}
\newacronym{glcm}{GLCM}{Gray Level Co-occurrence Matrix}
\newacronym{gt}{GT}{Ground-Truth}
\newacronym{VOD}{VOD}{Video On Demand}
\newacronym{HLS}{HLS}{HTTP Live Streaming}
\newacronym{CRF}{CRF}{Constant Rate Factor}
\newacronym{dash}{DASH}{Dynamic Adaptive Streaming over HTTP}
\newacronym{GP}{GP}{Gaussian Process}
\newacronym{TC}{TC}{Temporal Coherency}

%% file: figs/RD_firstPage.tex
\definecolor{s1}{RGB}{0, 153, 153}
\begin{tabular}{ll}

\begin{subfigure}{0.25 \textwidth}
\begin{tikzpicture}
\tikzstyle{every node}=[font=\tiny]
\begin{axis}
[ 
	width=\textwidth,, 
	height = 0.2 \textheight,
	xlabel= $\text{log}_{\text{2}}$ of bit/sec, 
	ylabel=Scaled-PSNR,
	xmax=18 , ymax=47,
    legend style={at={(axis cs:6.5,35)},anchor=south west}
] 
	\addplot [only marks, mark=o,draw=red, mark options={scale=0.5}] table[y=psnr_y_0,x=rate_0, mark=none]{figs/RD1_first_page.csv};
	\addplot [only marks, mark=o,draw=blue, mark options={scale=0.5}] table[y=psnr_y_1,x=rate_1, mark=none]{figs/RD1_first_page.csv};
	\addplot [only marks, mark=o,draw=green, mark options={scale=0.5}] table[y=psnr_y_2,x=rate_2, mark=none]{figs/RD1_first_page.csv};
	\addplot [only marks, mark=o,draw=orange, mark options={scale=0.5}] table[y=psnr_y_3,x=rate_3, mark=none]{figs/RD1_first_page.csv};
	
	\draw [dashed,color=black] (14.64745843,55) -- (14.64745843,0);
	\draw [dashed,color=black] (12.99859043,55) -- (12.99859043,0);
	\draw [dashed,color=black] (12.14911199,55) -- (12.1491119,0);
	

\legend{540p, 720p, 1080p, 2160p}
\end{axis} 
\end{tikzpicture}

\end{subfigure}
&
\begin{subfigure}{0.25 \textwidth}
\begin{tikzpicture}
\tikzstyle{every node}=[font=\tiny]
\begin{axis}
[ 
	width=\textwidth,, 
	height = 0.2 \textheight,
	xlabel= $\text{log}_{\text{2}}$ of bit/sec,
	legend style={at={(axis cs:13.5,29)},anchor=south west}, 
	xmax=18, ymax=47
] 
	\addplot [only marks, mark=o,draw=red, mark options={scale=0.5}] table[y=psnr_y_0,x=rate_0, mark=none]{figs/RD2_first_page.csv};
	\addplot [only marks, mark=o,draw=blue, mark options={scale=0.5}] table[y=psnr_y_1,x=rate_1, mark=none]{figs/RD2_first_page.csv};
	\addplot [only marks, mark=o,draw=green, mark options={scale=0.5}] table[y=psnr_y_2,x=rate_2, mark=none]{figs/RD2_first_page.csv};
	\addplot [only marks, mark=o,draw=orange, mark options={scale=0.5}] table[y=psnr_y_3,x=rate_3, mark=none]{figs/RD2_first_page.csv};
	
	\draw [dashed,color=black] (9.177419538,55) -- (9.177419538,0);
	\draw [dashed,color=black] (7.54689446,55) -- (7.54689446,0);
	\draw [dashed,color=black] (6.62935662,55) -- (6.62935662,0);

\legend{540p, 720p, 1080p, 2160p}
\end{axis} 

\end{tikzpicture}

\end{subfigure}

\end{tabular}

%% file: figs/RD_convex_hull_ladder.tex
\definecolor{s1}{RGB}{0, 153, 153}
\begin{tabular}{ll}

\begin{subfigure}{0.25 \textwidth}
\begin{tikzpicture}
\tikzstyle{every node}=[font=\tiny]
\begin{axis}
[ 
	width=\textwidth, 
	height = 0.2 \textheight,
	xlabel= $\text{log}_{\text{2}}$ of bit/sec, ylabel= Scaled PSNR,
	legend style={at={(0.9,-0.25)},legend columns=-1},
] 
	\addplot [only marks, mark=o,draw=red, mark options={scale=0.5}] table[y=psnr_y_0,x=rate_0, mark=none]{figs/RD2_first_page.csv};
	\addplot [only marks, mark=o,draw=blue, mark options={scale=0.5}] table[y=psnr_y_1,x=rate_1, mark=none]{figs/RD2_first_page.csv};
	\addplot [only marks, mark=o,draw=green, mark options={scale=0.5}] table[y=psnr_y_2,x=rate_2, mark=none]{figs/RD2_first_page.csv};
	\addplot [only marks, mark=o,draw=orange, mark options={scale=0.5}] table[y=psnr_y_3,x=rate_3, mark=none]{figs/RD2_first_page.csv};

\end{axis} 

\end{tikzpicture}

\caption{Full rate-quality points}
\end{subfigure}
&
\begin{subfigure}{0.25 \textwidth}
\begin{tikzpicture}
\tikzstyle{every node}=[font=\tiny]
\begin{axis}
[ 
	width=\textwidth, 
	height = 0.2 \textheight,
	xlabel= $\text{log}_{\text{2}}$ of bit/sec, ylabel= Scaled PSNR,
	legend style={at={(0.9,-0.25)},legend columns=-1},nodes near coords] 
	\addplot [mark=o,draw=red, mark options={scale=0.5}, point meta=explicit symbolic]
coordinates {
(6.374822142, 29.159032) [1]
(6.771918583, 30.217693) [1]
(7.136324705, 31.206483) [1]
(7.641546029, 32.661095) 
}; 

	\addplot [mark=o,draw=blue, mark options={scale=0.5}, point meta=explicit symbolic]
coordinates {
(7.641546029, 32.661095) [2]
(7.989976286, 33.708306) [2]
(8.327204559, 34.773269) [2]
(8.662953148, 35.950826)
}; 

	\addplot [mark=o,draw=green, mark options={scale=0.5}, point meta=explicit symbolic]
coordinates {
(8.662953148, 35.950826) [3]
(8.987889281, 37.160598) [3]
(9.301781962, 38.292469) [3]
(9.611264612, 39.307597) [3]
(9.923152895, 40.307246) [3]
(10.29739492, 41.369069)
};

	\addplot [mark=o,draw=orange, mark options={scale=0.5}, point meta=explicit symbolic]
coordinates {
(10.29739492, 41.369069) [4]
(10.60508129, 42.271006) [4]
(10.92226104, 43.096775) [4]
(11.29872704, 43.888078) [4]
(11.6895547 ,44.559775 ) [4]
(12.17764144, 45.172785) [4]
(12.75851004, 45.705919) [4]
(13.59612772, 46.182524) [4]
(14.58646878, 46.604931) [4]
(15.49169013, 47.101336) [4]
(16.41425464, 47.754936) [4]
};

\end{axis} 
\end{tikzpicture}

\caption{Convex hull}
\end{subfigure}

\\

\begin{subfigure}{0.25 \textwidth}
\begin{tikzpicture}
\tikzstyle{every node}=[font=\tiny]
\begin{axis}
[ 
	width=\textwidth, 
	height = 0.2 \textheight,
	xlabel= $\text{log}_{\text{2}}$ of bit/sec, ylabel= Scaled PSNR,
	legend style={at={(0.9,-0.25)},legend columns=-1},nodes near coords] 
	\addplot [mark=o,draw=red, mark options={scale=0.5}, point meta=explicit symbolic]
coordinates {
(6.374822142, 29.159032) 
(6.771918583, 30.217693) 
(7.136324705, 31.206483) 
(7.641546029, 32.661095) 
}; 

	\addplot [mark=o,draw=blue, mark options={scale=0.5}, point meta=explicit symbolic]
coordinates {
(7.641546029, 32.661095) 
(7.989976286, 33.708306) 
(8.327204559, 34.773269) 
(8.662953148, 35.950826)
}; 

	\addplot [mark=o,draw=green, mark options={scale=0.5}, point meta=explicit symbolic]
coordinates {
(8.662953148, 35.950826) 
(8.987889281, 37.160598) 
(9.301781962, 38.292469) 
(9.611264612, 39.307597) 
(9.923152895, 40.307246) 
(10.29739492, 41.369069)
};

	\addplot [mark=o,draw=orange, mark options={scale=0.5}, point meta=explicit symbolic]
coordinates {
(10.29739492, 41.369069) 
(10.60508129, 42.271006) 
(10.92226104, 43.096775) 
(11.29872704, 43.888078) 
(11.6895547 ,44.559775 ) 
(12.17764144, 45.172785) 
(12.75851004, 45.705919) 
(13.59612772, 46.182524) 
(14.58646878, 46.604931) 
(15.49169013, 47.101336) 
(16.41425464, 47.754936) 
};

\node[anchor=west] (source) at (axis cs:12,30){$P_v(s_1)$};
\node (destination) at (axis cs:7.2,30){};
\draw[->](source)--(destination);
\draw [dashed,color=black] (7.6,55) -- (7.6,0);

\node[anchor=west] (source) at (axis cs:12,32.5){$P_v(s_2)$};
\node (destination) at (axis cs:8.5,32.5){};
\draw[->](source)--(destination);
\draw [dashed,color=black] (8.6,55) -- (8.6,0);

\node[anchor=west] (source) at (axis cs:12,35){$P_v(s_3)$};
\node (destination) at (axis cs:10,35){};
\draw[->](source)--(destination);
\draw [dashed,color=black] (10.2,55) -- (10.2,0);

\end{axis} 
\end{tikzpicture}

\caption{Cross-points}
\end{subfigure}
&
\begin{subfigure}{0.25 \textwidth}
\begin{tikzpicture}
\tikzstyle{every node}=[font=\tiny]
\begin{axis}
[ 
	width=\textwidth, 
	height = 0.2 \textheight,
	xlabel= $\text{log}_{\text{2}}$ of bit/sec, ylabel= Scaled PSNR,
	legend style={at={(0.9,-0.25)},legend columns=-1},nodes near coords] 
	
	\addplot [mark=o,draw=red, mark options={scale=0.5}, point meta=explicit symbolic]
coordinates {
(6.5, 1)
(7.2, 1)
(7.2, 2)

}; 

\addplot [mark=o,draw=blue, mark options={scale=0.5}, point meta=explicit symbolic]
coordinates {

(7.2, 2)
(8.5, 2)
(8.5, 3)

}; 

	\addplot [mark=o,draw=green, mark options={scale=0.5}, point meta=explicit symbolic]
coordinates {

(8.5, 3)
(10, 3)
(10, 4)

}; 

\addplot [mark=o,draw=orange, mark options={scale=0.5}, point meta=explicit symbolic]
coordinates {

(10, 4)
(16, 4)

};

\end{axis} 
\end{tikzpicture}

\caption{Bitrate ladder}
\end{subfigure}

\end{tabular}

%% file: figs/algorithm.tex
\begin{algorithm}
\caption{Ensemble aggregator $Agr$}
\label{algorithm1}
\begin{algorithmic}
\STATE $\textbf{input: } \hat{L}^{Cl}_{v,S}, \hat{L}^{Rg}_{v,S}, \text{isFast, MinRate, MaxRate}$
\STATE $\textbf{output: } \hat{L}_{v,S}$

\FOR{ $r \coloneqq \text{MinRate \textbf{to} MaxRate}$}
    \STATE $\hat{i}^{Cl} \leftarrow L_{v,S}^{Cl}(r)$
    \STATE $\hat{i}^{Rg} \leftarrow L_{v,S}^{Rg}(r)$
    \IF{$\hat{i}^{Cl} = \hat{i}^{Rg}$}
    \STATE $i^*\leftarrow\hat{i}^{Cl} $
    \ELSE
    \IF{$\text{isFast}$}
    \STATE $i^* \leftarrow \argmaxA_i E(v;r,s_i) \text{ where }i \in \{\hat{i}^{Cl}, \hat{i}^{Rg}\} $
    \ELSE
    \STATE $i^* \leftarrow \argmaxA_i E(v;r,s_i) \text{ where } 1 \leq i \leq S $
    \ENDIF
    \ENDIF
    \STATE $\hat{L}_{v,S}(r) \leftarrow i^*$
\ENDFOR
\end{algorithmic}
\end{algorithm}

%% file: figs/SI_TI_scatter.tex
\definecolor{s1}{RGB}{0, 153, 153}
\centering
\tikzstyle{every node}=[font=\tiny]
\begin{tikzpicture}

	\begin{axis}[
	width=0.4\textwidth,, 
	height = 0.25 \textheight,
	xlabel= SI, 
	ylabel= TI,
	]
	\addplot [only marks, color=s1, mark options={scale=0.5}] table[x=avg_si,y=avg_ti]{figs/si_ti_scatter_info.csv};

\end{axis}
\end{tikzpicture}
\caption{The joint distribution of \gls{si} and \gls{ti}}

%% file: figs/average_res.tex
\renewcommand{\arraystretch}{1.15} 
\begin{tabular}{C{0.14\textwidth}|P{0.05\textwidth}P{0.1\textwidth}P{0.1\textwidth}}
\hline
Method & Accuracy   &  \gls{bd-br} vs. GT  &  \gls{bd-br} vs. static   \\
\hline	                  
Classification &  0.76 &  2.97\%  &  -11.45\%  \\ \hline
Regressor     &  0.83 &  1.37\%  &  -12.63\%  \\ \hline
Ensemble (fast)    &  0.90 &  0.89\%  &  -13.05\%  \\ \hline
Ensemble (full)    &  0.92 &  0.77\%  &  -13.14\%  \\ \hline

\end{tabular}

%% file: figs/hist_sample.tex
\definecolor{s1}{RGB}{0, 153, 153}
\begin{tabular}{cc}

\begin{subfigure}{0.25 \textwidth}
\begin{tikzpicture}

	\begin{axis}[ymin=0, xmin=-1, xmax=20,axis lines=left,xlabel={BD-BR (\%) w.r.t. GT ladder}, width = \textwidth,		height = 0.15 \textheight,   ylabel = {\# Sequences},    ybar=0pt, bar width=0.6, bar shift=0pt,ymax=75]
	\addplot[draw=none, fill=orange] table[x=bd,y=num]{figs/hist_classification_GT.csv};
\pgfplotsset{every x tick label/.append style={font=\tiny}, every y tick label/.append style={font=\tiny}, label style={font=\tiny}}

\end{axis}
\end{tikzpicture}
\end{subfigure}
&
\begin{subfigure}{0.25 \textwidth}
\begin{tikzpicture}

	\begin{axis}[ymin=0, xmin=-20, xmax=2,axis lines=left,xlabel={BD-BR (\%) w.r.t. static ladder}, width = \textwidth,		height = 0.15 \textheight,    ybar=0pt, bar width=0.6, bar shift=0pt,ymax=40]
	\addplot[draw=none, fill=s1] table[x=bd,y=num]{figs/hist_classification_static.csv};
\pgfplotsset{every x tick label/.append style={font=\tiny}, every y tick label/.append style={font=\tiny}, label style={font=\tiny}}

\end{axis}
\end{tikzpicture}
\end{subfigure}

\\ 
\multicolumn{2}{c}{(a) Classification}
\\
&

\\

\begin{subfigure}{0.25 \textwidth}
\begin{tikzpicture}

	\begin{axis}[ymin=0, xmin=-1, xmax=20,axis lines=left,xlabel={BD-BR (\%) w.r.t. GT ladder}, width = \textwidth,		height = 0.15 \textheight,   ylabel = {\# Sequences},    ybar=0pt, bar width=0.6, bar shift=0pt,ymax=75]
	\addplot[draw=none, fill=orange] table[x=bd,y=num]{figs/hist_regression_GT.csv};
\pgfplotsset{every x tick label/.append style={font=\tiny}, every y tick label/.append style={font=\tiny}, label style={font=\tiny}}

\end{axis}
\end{tikzpicture}
\end{subfigure}
&

\begin{subfigure}{0.25 \textwidth}
\begin{tikzpicture}

	\begin{axis}[ymin=0, xmin=-20, xmax=2,axis lines=left,xlabel={BD-BR (\%) w.r.t. static ladder}, width = \textwidth,		height = 0.15 \textheight,    ybar=0pt, bar width=0.6, bar shift=0pt,ymax=40]
	\addplot[draw=none, fill=s1] table[x=bd,y=num]{figs/hist_regression_static.csv};
\pgfplotsset{every x tick label/.append style={font=\tiny}, every y tick label/.append style={font=\tiny}, label style={font=\tiny}}

\end{axis}
\end{tikzpicture}
\end{subfigure}

\\ 
\multicolumn{2}{c}{(b) Regressor}
\\
&

\\
\begin{subfigure}{0.25 \textwidth}
\begin{tikzpicture}

	\begin{axis}[ymin=0, xmin=-1, xmax=20,axis lines=left,xlabel={BD-BR (\%) w.r.t. GT ladder}, width = \textwidth,		height = 0.15 \textheight,   ylabel = {\# Sequences},    ybar=0pt, bar width=0.6, bar shift=0pt,ymax=75]
	\addplot[draw=none, fill=orange] table[x=bd,y=num]{figs/hist_Ensemple_2enc_GT.csv};
\pgfplotsset{every x tick label/.append style={font=\tiny}, every y tick label/.append style={font=\tiny}, label style={font=\tiny}}

\end{axis}
\end{tikzpicture}
\end{subfigure}

&

\begin{subfigure}{0.25 \textwidth}
\begin{tikzpicture}

	\begin{axis}[ymin=0, xmin=-20, xmax=2,axis lines=left,xlabel={BD-BR (\%) w.r.t. static ladder}, width = \textwidth,		height = 0.15 \textheight,    ybar=0pt, bar width=0.56, bar shift=0pt,ymax=40]
	\addplot[draw=none, fill=s1] table[x=bd,y=num]{figs/hist_Ensemple_2enc_static.csv};
\pgfplotsset{every x tick label/.append style={font=\tiny}, every y tick label/.append style={font=\tiny}, label style={font=\tiny}}

\end{axis}
\end{tikzpicture}
\end{subfigure}

\\ 
\multicolumn{2}{c}{(c) Ensemble (fast)}
\\
&

\\
\begin{subfigure}{0.25 \textwidth}
\begin{tikzpicture}

	\begin{axis}[ymin=0, xmin=-1, xmax=20,axis lines=left,xlabel={BD-BR (\%) w.r.t. GT ladder}, width = \textwidth,		height = 0.15 \textheight,   ylabel = {\# Sequences},    ybar=0pt, bar width=0.6, bar shift=0pt,ymax=75]
	\addplot[draw=none, fill=orange] table[x=bd,y=num]{figs/hist_Ensemple_4enc_GT.csv};
\pgfplotsset{every x tick label/.append style={font=\tiny}, every y tick label/.append style={font=\tiny}, label style={font=\tiny}}

\end{axis}
\end{tikzpicture}
\end{subfigure}

&

\begin{subfigure}{0.25 \textwidth}
\begin{tikzpicture}

	\begin{axis}[ymin=0, xmin=-20, xmax=2,axis lines=left,xlabel={BD-BR (\%) w.r.t. static ladder}, width = \textwidth,		height = 0.15 \textheight,    ybar=0pt, bar width=0.6, bar shift=0pt,ymax=40]
	\addplot[draw=none, fill=s1] table[x=bd,y=num]{figs/hist_Ensemple_4enc_static.csv};
\pgfplotsset{every x tick label/.append style={font=\tiny}, every y tick label/.append style={font=\tiny}, label style={font=\tiny}}

\end{axis}
\end{tikzpicture}
\end{subfigure}
\\ 
\multicolumn{2}{c}{(d) Ensemble (full)}

\end{tabular}

%% file: figs/comp_vs_bdbr.tex
\definecolor{s1}{RGB}{0, 153, 153}
\begin{tabular}{c}

\begin{subfigure}{0.5 \textwidth}
\begin{tikzpicture}
\tikzstyle{every node}=[font=\tiny]

\begin{axis}
[ 
	grid=both,
	width=\textwidth, 
	height = 0.2 \textheight,
	xlabel= Encoding time (sec) needed for ladder computation, ylabel= BD-BR (\%) w.r.t. static ladder,
	legend style={at={(0.9,-0.25)},legend columns=-1},nodes near coords] 
	\addplot [mark=o,draw=red, mark options={scale=0.5}, point meta=explicit symbolic]
coordinates {
(0,          -11.4547936291486)
(0,          -12.7371173009483)
(945465.41,  -13.0464499937583)
(1604321.92, -13.1354785596268)
(11266663.15,-13.37692478766)   
}; 
\node[anchor=west] at (axis cs:0,-11.5){Classification (0\%)};
\node[anchor=west] at (axis cs:100,-12.7){Regression (0\%)};
\node[anchor=west] at (axis cs:900000,-12.95){Ensemble fast (8\%)};
\node[anchor=west] at (axis cs:1400000,-13.3){Ensemble full (14\%)};
\node[anchor=west] at (axis cs:9800000,-13.2){Ground truth (100\%)};

\end{axis} 
\end{tikzpicture}

\end{subfigure}

\end{tabular}